\documentclass{article}
\usepackage[T1]{fontenc}
\usepackage{a4wide}
\usepackage{graphics}
\usepackage{subfigure}

\makeatletter

\providecommand{\LyX}{L\kern-.1667em\lower.25em\hbox{Y}\kern-.125emX\@}
\newlength{\LyXMinipageIndent}
\let\SF@@footnote\footnote
\def\footnote{\ifx\protect\@typeset@protect
    \expandafter\SF@@footnote
  \else
    \expandafter\SF@gobble@opt
  \fi
}
\expandafter\def\csname SF@gobble@opt \endcsname{\@ifnextchar[%]
  \SF@gobble@twobracket
  \@gobble
}
\edef\SF@gobble@opt{\noexpand\protect
  \expandafter\noexpand\csname SF@gobble@opt \endcsname}
\def\SF@gobble@twobracket[#1]#2{}

\usepackage{verbatim}

\usepackage{subfigure}
\usepackage{graphicx}
\usepackage{epsfig}
\usepackage{acronym}
\newcommand{\ee}{e^{+}e^{-}}
\newcommand{\pipipi}{\pi ^{+}\pi ^{-}\pi ^{0}}
\newcommand{\eeg}{e^{+}e^{-}(\gamma )}
\newcommand{\pipig}{\pi ^{+}\pi ^{-}(\gamma )}
\newcommand{\mumug}{\mu ^{+}\mu ^{-}(\gamma )}

\newcommand{\omgee}{\ensuremath 0.595 \pm 0.014 \pm 0.009}
\newcommand{\omgeebr}{\ensuremath 0.528 \pm 0.012 \pm 0.007}

\newcommand{\omcs}{\ensuremath 1457 \pm 23 \pm 19}
\newcommand{\ommass}{\ensuremath 782.71 \pm 0.07 \pm 0.04}
\newcommand{\omwidth}{\ensuremath 8.68 \pm 0.23 \pm 0.10}

\makeatother

\begin{document}

\acrodef{RDM}{resonant depolarization method}

\thispagestyle{headings}

\title{Measurement of \protect\( \omega \protect \) meson parameters in \protect\( \pipipi \protect \)
decay mode with CMD-2 \thanks{
Work is supported in part by grants RFBR-98-02-17851, RFBR-99-02-17053, RFBR-99-02-17119 
}}

\date{~}

\author{R.R.~Akhmetshin, E.V.~Anashkin\thanks{
contact person. e-mail: E.V.Anashkin@inp.nsk.su
}, V.M.~Aulchenko,  V.Sh.~Banzarov, L.M.~Barkov, \and S.E.~Baru,
N.S.~Bashtovoy, A.E.~Bondar, D.V.~Bondarev, D.V.~Chernyak,
 \and S.I.~Eidelman,
 G.V.~Fedotovich, N.I.~Gabyshev, A.A.~Grebeniuk, D.N.~Grigoriev,
 \and B.I.~Khazin, I.A.~Koop, L.M.~Kurdadze, A.S.~Kuzmin, I.B.~Logashenko,
 \and P.A.~Lukin, A.P.~Lysenko,
I.N.~Nesterenko,  V.S.~Okhapkin, E.A.~Perevedentsev,  \and A.A.~Polunin,
 T.A.~Purlatz, N.I.~Root, A.A.~Ruban,
N.M.~Ryskulov,  \and A.G.~Shamov,  Yu.M.~Shatunov, A.I.~Shekhtman, A.E.~Sher,
B.A.Shwartz,  \and V.A.~Sidorov, A.N.~Skrinsky, V.P.~Smakhtin, I.G.~Snopkov,
E.P.~Solodov,  \and P.Yu.~Stepanov, A.I.~Sukhanov, V.M.~Titov, A.A.~Valishev,
\and Yu.V.~Yudin, S.G.~Zverev \\
 \vspace{2mm} \textit{Budker Institute of Nuclear Physics, Novosibirsk, 630090,
Russia}\\
  \vspace{2mm} J.A.~Thompson\\
 \vspace{2mm} \textit{University of Pittsburgh, Pittsburgh, PA 15260, USA}\\
\vspace{2mm} S.K.~Dhawan, V.W.~Hughes\\
 \vspace{2mm} \textit{Yale University, New Heaven, CT 06511, USA}}

\maketitle
\begin{abstract}
About 11 200 \( \ee \to \omega \to \pipipi  \) events selected in the
center of mass energy range  from 760 to 810 MeV were used for the 
measurement of the \( \omega  \) meson parameters. The following results 
have been obtained: \( \sigma _{0}=(\omcs ) \)
nb, \( m_{\omega }=(\ommass ) \) MeV/c\( ^{2} \), 
\( \Gamma _{\omega }=(\omwidth ) \) MeV, 
\( \Gamma _{\ee }\cdot  \)Br\( (\omega \rightarrow \pipipi )=
(\omgeebr) \cdot 10^{-3} \) MeV. 
\end{abstract}

\section{Introduction}

High precision measurements of the \( \omega  \) meson parameters 
provide valuable information for testing various theoretical models describing 
interactions of light quarks. This paper presents a precise 
determination of the mass, total width and
leptonic width of the \( \omega  \), based on its dominant decay mode, 
\( \omega \rightarrow \pipipi  \).

The data sample was collected with the CMD-2 detector in 1994-1995
while scanning the center of mass energy range 2\( E_{beam} \)
from 760 to 810 MeV at the high luminosity collider VEPP-2M~\cite{VEPP}.
The \acl{RDM}~\cite{Dep} was used for the precise calibration of the
beam energy at each point. The integrated luminosity of 141 nb\( ^{-1} \) 
corresponds to \( \sim 7\times 10^{4} \) \( \omega  \) meson decays.

\section{CMD-2 detector}

The CMD-2 detector has been described in detail elsewhere~\cite{CMD2}. It is
a general purpose detector consisting of a drift chamber (DC) and proportional
Z-chamber (ZC), both inside a thin (0.38 \( X_{0} \))
superconducting solenoid with a field of 1~T. 

The barrel calorimeter placed outside the solenoid consists of 892 CsI crystals
of 6\( \times  \)6\( \times  \)15~cm\( ^{3} \) size. It covers polar angles
from 0.8 to 2.3 radian. The energy resolution for photons 
%in the CsI calorimeter 
is about 8\% in the energy range from 100 to 700~MeV.

The trigger signal is generated either by the charged trigger based on DC and
ZC hits~\cite{TF} or by the neutral trigger~\cite{NT} which takes into account
the number 
and relative position 
of clusters detected in the CsI calorimeter as well as the total
energy deposition. These two independent triggers have been used 
to study the trigger efficiency.

\section{Analysis}

%The \( \omega  \) meson parameters were measured by the  
%\( \omega \rightarrow \pi ^{+}\pi ^{-}\pi ^{0} \) decay mode.

Events with two tracks originating from the same vertex, 
each with a polar angle \( 0.85<\theta <\pi -0.85 \) within 
the fiducial volume of the detector, were selected for further analysis.

\begin{figure}[!htb]
{\centering \begin{tabular}{cc}
\subfigure[~]{\resizebox*{0.35\textwidth}{!}{\includegraphics{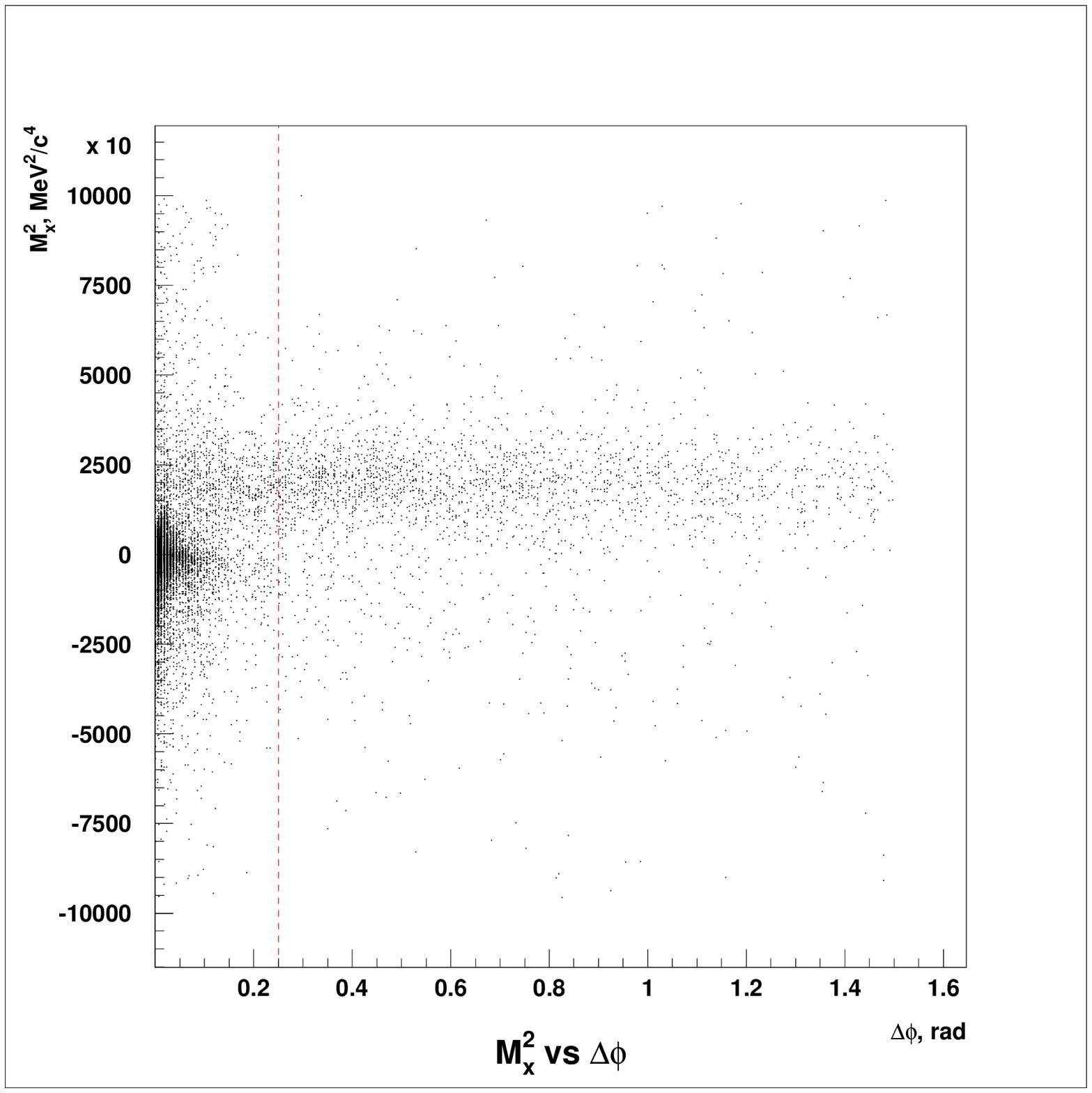}}} &
\subfigure[~]{\resizebox*{0.35\textwidth}{!}{\includegraphics{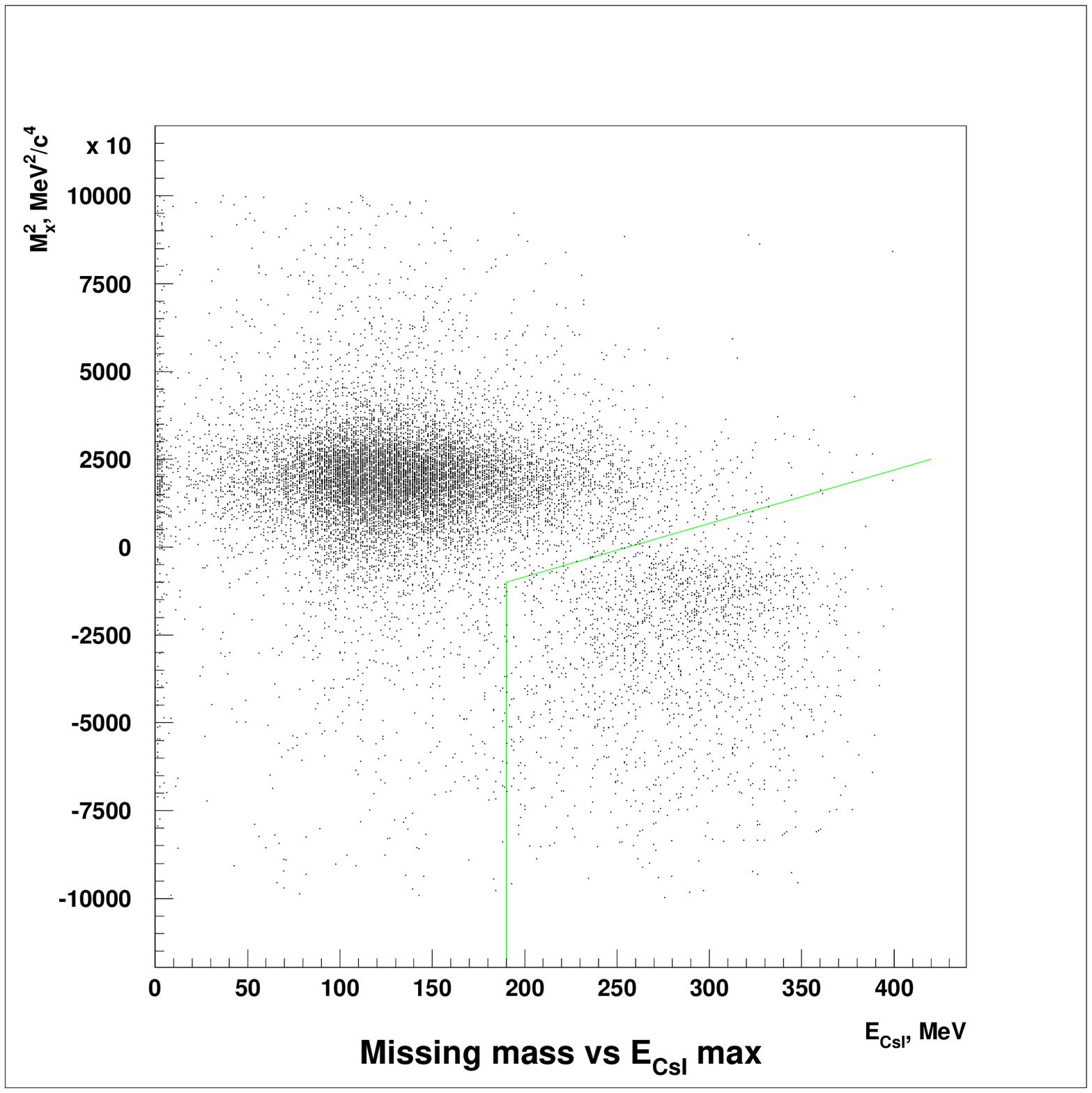}}} \\
\end{tabular}\par}

\caption{\label{fig:bg_cuts}Graphic presentation of cuts on 
\protect\( \Delta \phi \protect \) (a) and 
\protect\( M^{2}_{X}\protect \) vs \protect\( E^{max}_{CsI}\protect \) (b).
(b) contains events after the cut (a). 
In Fig. (b) the lower right corner corresponds to rejected events.}
\end{figure}

To minimize a systematic error of the detection efficiency,
only DC information has been used for the selection of 
$\omega \rightarrow \pi^{+} \pi^{-} \pi^{0}$ events.
Most of the background comes from the processes with the hard photon 
emission:
\[
e^{+}e^{-}\rightarrow e^{+}e^{-}\gamma ,\, 
\pi ^{+}\pi ^{-}\gamma ,\, \mu ^{+}\mu ^{-}\gamma .\]
 These processes have the same signature as the reaction 
$e^{+}e^{-}\rightarrow \pi^{+}\pi^{-}\pi^{0}$,
except for the very different acollinearity angle 
(\( \Delta \phi =\pi -|\varphi _{1}-\varphi _{2}| \))
distribution peaked near \( \Delta \phi =0 \). Thus, the rejection of events
with a small \( \Delta \phi  \) drastically reduces the background, at the
same time decreasing the number of $\pi^+\pi^-\pi^0$ events. 
The value of 
$\Delta \phi =0.25$ was used as a reasonable compromise (see 
Fig.~\ref{fig:bg_cuts}-a). 

Additional background suppression was achieved using the 
"$\pi^{+}\pi^{-}$ missing mass" parameter $M_{X}$ 
which is calculated assuming charged particles to be 
pions and taking into account energy-momentum conservation. For real 
\( \pi ^{+}\pi ^{-}\pi ^{0} \)
events the distribution of the missing mass squared has a peak in the region
of \( M^{2}_{\pi ^{0}} \) in contrast to the background processes which have a
peak around zero for \( e^{+}e^{-}\rightarrow \pipig ,\, \mumug  \) or in the
negative region for \( e^{+}e^{-}\rightarrow \eeg  \). 
Further rejection of events of the process $e^{+}e^{-}\rightarrow e^+e^-\gamma$
which has the largest cross section among the backround processes, 
is based on the maximum energy deposition of two charged particles in 
the calorimeter $E_{CsI}^{max}$. The corresponding cuts shown by two lines
in Fig.~\ref{fig:bg_cuts}-b reject its lower right corner mostly
populated by events of this background source.
%Figure~\ref{fig:bg_cuts}-b shows the squared missing mass of two charged 
%particles versus the maximum energy deposition of these two particles in 
%the calorimeter. The lines show the cut applied for the separation of 
%\( \pi ^{+}\pi ^{-}\pi ^{0} \) events from the background.

The number of \( \pi^{+}\pi^{-}\pi^{0} \) events was obtained in
two different ways. The first method was to fit 
%the histograms of the  \( \pi ^{+}\pi ^{-} \) missing mass 
the $M_{X}$ distributions
with the sum of Gaussian functions 
describing \( \pi ^{+}\pi ^{-}\pi ^{0} \) and background events. In the second 
method the cosmic and beam background was rejected by fitting the 
distribution of the \( z \)-coordinate of the vertex with the sum of 
a Gaussian function and a  constant background. In the last case 
the  remaining  \( \ee \to \eeg ,\, \pipig ,\, \mumug  \) events
were simulated and subtracted from the total number of events at each point
according to the corresponding integrated luminosity. Both approaches 
gave the same result within  statistical errors.

At each energy the  \( \pipipi  \) production cross section
was calculated
according to the formula: 
\begin{eqnarray*}
\sigma (e^{+}e^{-}\rightarrow \pipipi )=\frac{N_{\pipipi }}
{L\cdot \varepsilon _{trig}\cdot \varepsilon _{MC}
\cdot \varepsilon _{M^{2}_{X}}\cdot (1+\delta _{rad})}\, ,
%\cdot (1+\delta _{E})}\, ,
\end{eqnarray*}
where \( N_{\pipipi } \) is the number of events; \( L \) is the integrated
luminosity determined from large angle Bhabha events with the help of 
the procedure described in~\cite{Prep99}; \( \delta _{rad} \) is the 
radiative correction calculated according to~\cite{Rad} with 
an accuracy better than 0.5\%; and  \( \varepsilon _{trig},\, \varepsilon _{MC},\, 
\varepsilon _{M^{2}_{X}} \)
are respectively the trigger efficiency, 
 the geometrical efficiency (acceptance) multiplied by 
the reconstruction efficiency, and the efficiency of the cut shown in 
Fig.~\ref{fig:bg_cuts}-b.
The acceptance is the probability to detect two pions from the \( \omega  \)
decay within a given solid angle. It was
calculated by Monte Carlo simulation taking into 
account radiative photons emitted by initial electrons.
% which allow correctly calculate registration
% efficiency for applied cuts at each energy point. 

\begin{comment}
This efficiency depends on the beam energy and is slightly higher (about 2.5\%)
at the right slope of the resonance.
\end{comment}
The efficiencies \( \varepsilon _{MC} \) and \( \varepsilon _{M^{2}_{X}} \) 
were calculated by Monte Carlo simulation. Their systematic errors were 
estimated with the help of special \char`\"{}test\char`\"{} events 
obtained as a result of the constrained fit based on the information 
from the ZC and CsI calorimeter only. About 40\% of the \( \pipipi  \) 
events have two clusters in the CsI calorimeter resulting from a neutral 
pion decay. Using the polar and azimuthal angles of these clusters as well 
as the hits of charged tracks in the ZC, one can reconstruct the 
\( \omega \rightarrow \pipipi  \) event without DC information. ``Test'' 
events with  the neutral trigger were also used to determine the 
charged trigger efficiency.

Typical values of the efficiencies and corrections are presented 
in Table~\ref{tab:eff} for \( 2E_{beam}=782.0 \)~MeV 
(the \( \omega  \) meson peak). 
\begin{table}[!htb]

\caption{\label{tab:eff}Efficiencies, corrections and their errors at 
\protect\( 2E_{beam}=782.0\protect \)~MeV}
\medskip{}
{\centering \begin{tabular}{cccc}
\hline 
Efficiency &
 Value, \%&
 Stat. error, \%&
 Syst. error, \%\\
\hline 
\( \varepsilon _{MC} \)&
 19.0 &
 0.1 &
 0.1 \\
\( \varepsilon _{trig} \)&
 99.5 &
 0.2 &
 0.1 \\
\( \varepsilon _{M^{2}_{X}} \)&
 99.2 &
 0.2 &
 0.2 \\
1+\( \delta _{rad} \)&
 78.5 &
0.1 &
 0.5 \\
\hline 
\end{tabular}\par}\end{table}

%The beam energy at each point was measured by the \acl{RDM}~\cite{Dep}. 
The integrated luminosity, radiative correction, 
number of selected $\pi^+\pi^-\pi^0$ events and cross section for 
\( \ee \to \omega \to \pipipi  \) at each energy are presented in 
Table~\ref{t:omega}.

\begin{table}[!htb]

\caption{\label{t:omega}Integrated luminosity, radiative correction, 
number of events and cross section for 
\protect\( \ee \to \pipipi \protect \)}\medskip{}

{\centering \begin{tabular}{ccccc}
\hline 
\( E_{beam} \), MeV&
\( \int Ldt \), nb\( ^{-1} \)&
\( \delta _{rad} \)&
\( N_{\pipipi } \)&
\( \sigma (\omega \to \pipipi ) \), nb\\
\hline 
380.092&
6.20\( \pm  \)0.10&
-0.183&
64\( \pm  \)11&
67\( \pm  \)11\\
382.083&
10.79\( \pm  \)0.14&
-0.191&
115\( \pm  \)13&
69\( \pm  \)8\\
385.053&
8.23\( \pm  \)0.12&
-0.206&
216\( \pm  \)17&
175\( \pm  \)15\\
387.190&
6.50\( \pm  \)0.11&
-0.220&
263\( \pm  \)18&
272\( \pm  \)21\\
389.087&
6.62\( \pm  \)0.11&
-0.232&
739\( \pm  \)29&
771\( \pm  \)40\\
390.087&
7.03\( \pm  \)0.11&
-0.232&
1155\( \pm  \)35&
1165\( \pm  \)50\\
391.113&
19.03\( \pm  \)0.18&
-0.215&
4080\( \pm  \)65&
1455\( \pm  \)30\\
392.119&
10.43\( \pm  \)0.14&
-0.172&
2104\( \pm  \)47&
1306\( \pm  \)44\\
393.018&
5.17\( \pm  \)0.09&
-0.116&
753\( \pm  \)28&
882\( \pm  \)43\\
395.047&
9.31\( \pm  \)0.12&
0.031&
747\( \pm  \)29&
407\( \pm  \)19\\
397.068&
9.17\( \pm  \)0.08&
0.178&
403\( \pm  \)22&
192\( \pm  \)11\\
400.000&
9.75\( \pm  \)0.12&
0.358&
313\( \pm  \)19&
124\( \pm  \)8\\
405.071&
14.29\( \pm  \)0.15&
0.613&
244\( \pm  \)18&
55\( \pm  \)4\\
\hline 
\end{tabular}\par}\end{table}

\section{\protect\( \omega \protect \) meson parameters}

The experimental data were fitted with a function which includes 
the interference
of the \( \omega  \) and \( \phi  \) mesons and non-resonant background:

\begin{equation}
\label{eq:cs}
\sigma _{3\pi }(s)=\frac{F_{3\pi }(s)}{s^{3/2}}\cdot |A_{\omega }+e^{i\alpha }A_{\phi }+A_{bg}|^{2},
\end{equation}
\[
A_{V}=\frac{m_{V}^{2}\Gamma _{V}\sqrt{\sigma _{V}m_{V}/F_{3\pi }(m_{V}^{2})}}{s-m_{V}^{2}+i\sqrt{s}\Gamma _{V}(s)},\]

\[
A_{bg}=m_{\omega }^{3/2}\sqrt{\sigma _{bg}/F_{3\pi }(m_{\omega }^{2})}\, ,\]

\[
\Gamma _{\omega }(s)=\Gamma _{\omega }\cdot \Bigl (Br_{\pi ^{+}\pi ^{-}}\frac{m^{2}_{\omega }F_{2\pi }(s)}{sF_{2\pi }(m^{2}_{\omega })}+Br_{\pi ^{0}\gamma }\frac{F_{\pi ^{0}\gamma }(s)}{F_{\pi ^{0}\gamma }(m^{2}_{\omega })}+Br_{3\pi }\frac{\sqrt{s}F_{3\pi }(s)}{m_{\omega }F_{3\pi }(m^{2}_{\omega })}\Bigr ),\]

\[
F_{\pi ^{0}\gamma }(s)=(\sqrt{s}(1-m_{\pi ^{0}}^{2}/s))^{3},\, F_{2\pi }(s)=(s/4-m_{\pi }^{2})^{3/2},\]
 
\noindent
where \( m_{V},\, \Gamma _{V},\, \sigma _{V} \) are mass, width and peak cross
section (\( s=m_{V}^{2} \)) for the vector meson \( \omega  \) or \( \phi  \);
\( \alpha  \) is a relative phase of \( \omega -\phi  \) mixing taken to be
\( (155\pm 15)^{\circ } \) according to~\cite{ND91}.  \( F_{3\pi }(s) \)
is a smooth function which describes the dynamics of 
\( V\to \rho \pi \to \pipipi  \) decay 
%\( V\to \pipipi  \) decay
including the phase space~\cite{KS}.
%\( F_{3\pi }(s) \) 
% was numerically calculated assuming the model
% \( V\to \rho \pi \to \pipipi  \). 
\( \Gamma _{\phi }(s) \) has
been parametrized similarly  to \( \omega  \) using 
the corresponding branching ratios and phase space factors~\cite{phi-3pi}.

The cross section values were fit by the function~(\ref{eq:cs}). 
The \( \omega  \) meson mass, width, peak cross section and background 
cross section were optimized, while the \( \phi  \) meson parameters 
were fixed at their world average values~\cite{pdg}. 

\begin{figure}[!htb]
{\setlength\parindent{0pt}
\begin{minipage}[t]{0.50\columnwidth}
\setlength\parindent{\LyXMinipageIndent}
\resizebox*{0.9\textwidth}{!}{\includegraphics{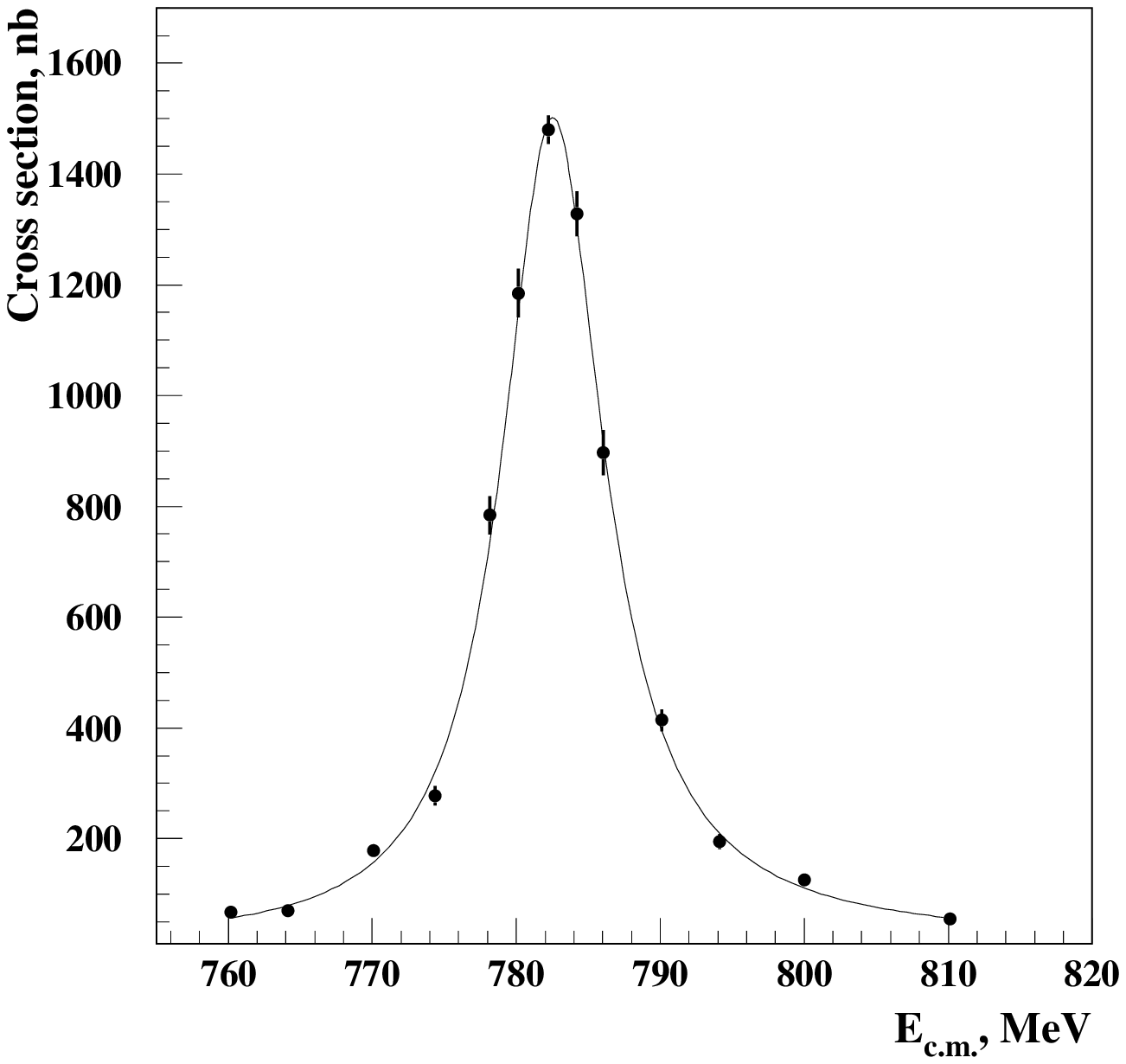}} 

\end{minipage}
}
\hfill{}
{\setlength\parindent{0pt}
\begin{minipage}[t]{0.50\columnwidth}
\setlength\parindent{\LyXMinipageIndent}
\resizebox*{0.9\textwidth}{!}{\includegraphics{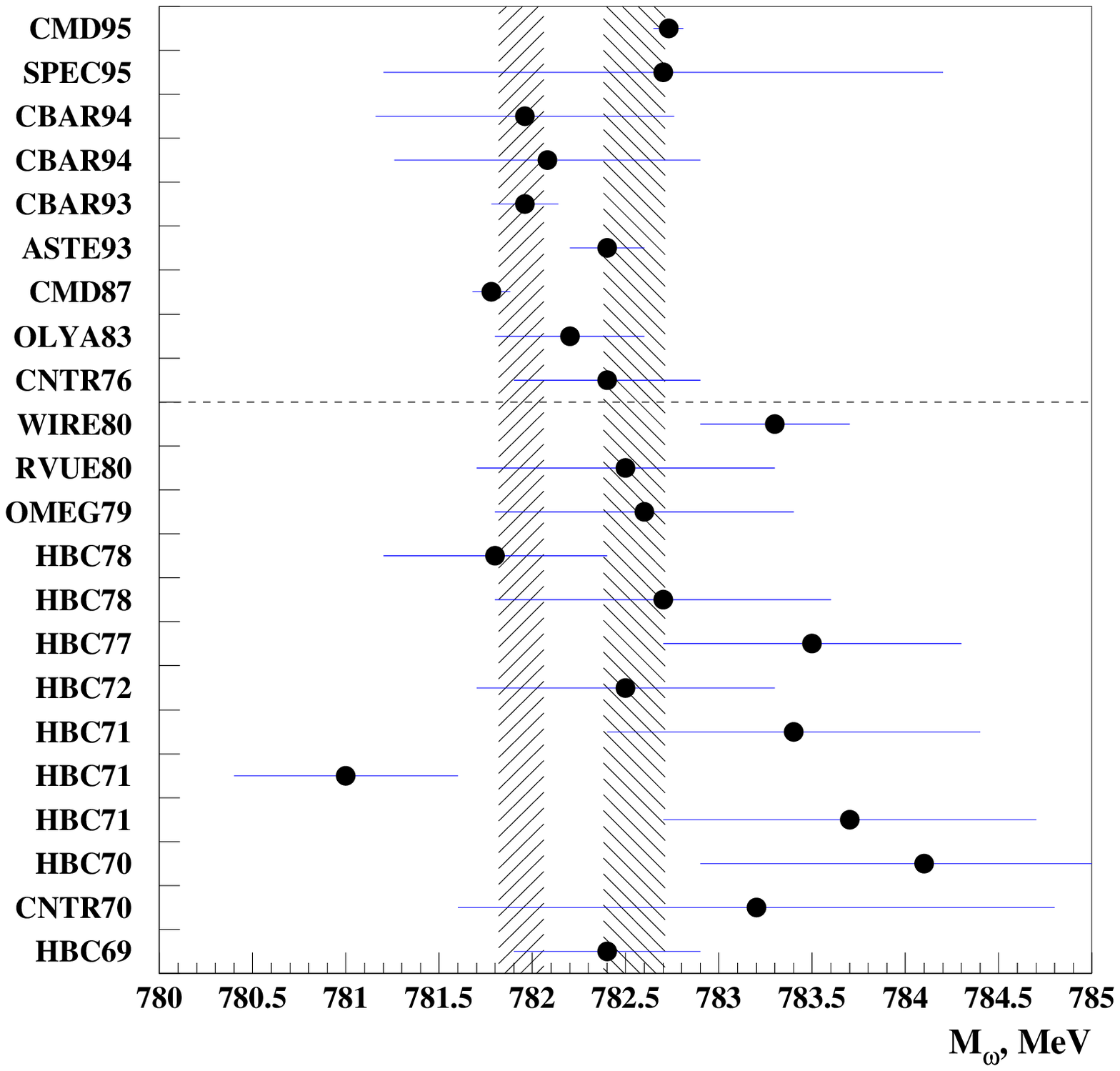}} 

\end{minipage}
}
\hfill{}
{\setlength\parindent{0pt}
\begin{minipage}[t]{0.50\columnwidth}
\setlength\parindent{\LyXMinipageIndent}

\caption{\label{fig:omega_fit}\protect\( \omega \protect \) meson excitation curve}

\end{minipage}
}
\hfill{}
{\setlength\parindent{0pt}
\begin{minipage}[t]{0.50\columnwidth}
\setlength\parindent{\LyXMinipageIndent}

\caption{\label{fig:om_mass}Experimental data on the
\protect\( \omega \protect \) meson mass. The left shaded bar corresponds 
to the current world average~\cite{pdg},
the right one --- to the world average before the CMD87 experiment. 
This work (CMD95) and experiments below the dashed line are not used for 
the current world average.}
\end{minipage}
}
\medskip

\end{figure}

The energy dependence of the cross section is shown 
in Fig.~\ref{fig:omega_fit} (experimental points and the optimal
fitting curve). 
 The following \( \omega  \) meson parameters were obtained from the fit:\\
$\sigma _{0}=(1457 \pm 23)$ nb, 
$M_{\omega}=(782.71 \pm 0.07)$ MeV/c$^{2}$,
$\Gamma_{\omega}=(8.68 \pm 0.23)$ MeV, 
$\sigma_{bg}=(12 \pm 5)$ nb.

{\setlength\parindent{0pt}
\begin{minipage}[t]{1.00\columnwidth}
\setlength\parindent{\LyXMinipageIndent}

The systematic error of \( \sigma _{0} \) is about 1.3\% and comes
from the following sources: \\

\noindent reconstruction efficiency \hfill{} 0.5\% ; \\
 trigger efficiency \hfill{} 0.1\% ;\\
 radiative corrections for the process \( e^{+}e^{-}\rightarrow \pipipi  \) 
\hfill{} 0.5\% ; \\
 decays in flight \hfill{} 0.1\% ;\\
 pion nuclear interaction \hfill{} 0.2\% ;\\
 solid angle uncertainty \hfill{} 0.3\% ; \\
luminosity determination \hfill{} 1.0\% . \\

\end{minipage}
}
\medskip

The systematic error of the mass was found to be about 40 keV
dominated by the stability of the beam energy. 

The systematic error of the width was found to be about 100 keV
dominated by the scatter of results of various fits corresponding to
different selection criteria.
   
\section{Discussion}

The measurements of \( \sigma _{0}(\omega \rightarrow \pipipi ) \) have been
performed by a number of groups from Orsay and Novosibirsk with the 
results presented in Table~\ref{sigma0}. 
\begin{table*}[!htb]

\caption{\label{sigma0}Results of \protect\( \sigma _{0}\protect \) 
measurements by various groups}
\medskip{}
{\centering \begin{tabular}{ccc}
\hline 
Group &
 \( \sigma _{0}(e^{+}e^{-}\rightarrow \omega \rightarrow \pipipi ) \), nb &
 Reference \\
\hline
 OSPK1,1969 &
 1590 \( \pm  \) 165 &
 \cite{ACO1}\\
 OSPK2,1972 &
 1800 \( \pm  \) 200 &
 \cite{ACO2}\\
 DM1,1980 &
 1410 \( \pm  \) 130 &
 \cite{DM1}\\
 OLYA,1982 &
 1390 \( \pm  \) 100 &
 \cite{OLYA82}\\
 OLYA,1984 &
 1420 \( \pm  \) 100 &
 \cite{OLYA84}\\
CMD, 1987 &
 1549 \( \pm  \) 57 &
 \cite{CMD87}\\
 ND, 1989 &
 1530 \( \pm  \) 77&
 \cite{ND91}\\
CMD-2 &
 1457 \( \pm \) 30 &
 This work  \\
\hline 
\end{tabular}\par}\bigskip{}
\end{table*}
 One can see that the result of this work 
$\sigma_{0}(\omega \rightarrow \pi^+\pi^-\pi^0) = (1457 \pm 30)$ 
nb does not contradict these measurements and is the most precise.

The cross section in the peak obtained in our experiment is related to 
the product 
\( \Gamma _{e^{+}e^{-}}\cdot  \)Br\( (\omega \rightarrow \pipipi ) \). 
To obtain this value, the fit with this product as a free parameter 
has been performed with the following result: 
\begin{eqnarray*}
\Gamma _{e^{+}e^{-}}\cdot \mbox {Br}(\omega \rightarrow \pipipi )=(\omgeebr )\, \mbox {keV}\, , & 
\end{eqnarray*}
 which is the most precise direct measurement of this quantity. Using 
\( \Gamma _{e^{+}e^{-}} \) from other experiments, one can obtain 
\( Br(\omega \rightarrow \pipipi ) \).
% according to the following formula: 
%\[
%Br(\omega \rightarrow \pi ^{+}\pi ^{-}\pi ^{0})=\sigma _{0}(\omega \rightarrow \pi ^{+}\pi ^{-}\pi ^{0})\cdot \frac{\Gamma _{\omega }}{\Gamma _{e^{+}e^{-}}}\cdot \frac{M^{2}_{\omega }}{12\pi }.\]
For example, for \( \Gamma _{e^{+}e^{-}}=(0.60\pm 0.02) \)~keV from~\cite{pdg},
\( Br(\omega \rightarrow \pipipi )=0.880\pm 0.020\pm 0.032 \) can be obtained.
Alternatively, taking \( Br(\omega \rightarrow \pipipi ) \) from other 
measurements,
\( \Gamma _{e^{+}e^{-}} \) can be calculated. 
For \( Br(\omega \rightarrow \pipipi )=0.888\pm 0.007 \)
(from~\cite{pdg}), we obtain for the leptonic width 
\( \Gamma _{\ee }=(\omgee ) \) keV or for the leptonic branching ratio
$\Gamma_{e^{+}e^{-}}/\Gamma_{\omega} = 
(6.85 \pm 0.11 \pm 0.11) \cdot 10^{-5}$.  
%Figure~\ref{fig:om_mass} shows the results of previous measurements of the 
%\( \omega  \) meson mass. CMD95 denotes the result of this work. 
%The left shaded bar corresponds to the current world average. This value 
%is dominated  by the CMD87 experiment with the reported precision 
%far better than in all other experiments. 

In Fig.~\ref{fig:om_mass} the result of this work (CMD95) is compared to
the previous measurements of the \( \omega  \) meson mass.  
The left shaded bar corresponds to the current world average. Its value 
is dominated  by the CMD87 experiment which was also performed at the
VEPP-2M collider with the CMD detector \cite{CMD87}. The reported precision
of CMD87 was much better than in all other experiments. Our new measurement 
gives the $\omega$ meson mass value 930 keV higher (more than seven 
standard deviations) than in CMD87. 

Since both measurements were performed at VEPP-2M and used the resonant 
depolarization method (RDM), thorough comparison of two experiments
has been carried out. 

We now assume that the difference between the two results is due to the fact
that \ac{RDM} measurements in CMD87 
were regularly performed at some side band resonance. Such a 
resonance could arise from a parasitic modulation of the depolarizer 
frequency since  
the RF device used for the \ac{RDM} had the power about $10^4$ -- $10^5$ times 
higher than required. Thus, the 
absolute calibration of the beam energy gave wrong results.

Unfortunately, after a lapse 
of more than ten years, it is impossible to reproduce the CMD87 environment
and prove the above hypothesis. However, we know that because of various 
technical problems  
inadequate attention was given at that time to the possibility of the  
low modulation leading to a side band resonance. 

Since the time of the CMD87 experiment, the  VEPP-2M collider and the 
\ac{RDM} hardware have been upgraded.
The applied RF power is of the order of a few mW excluding any ``parasitic'' 
depolarization. Furthermore, the  frequency spectrum of the depolarizer 
was investigated before the \ac{RDM} measurements and we believe that 
in the present experiment 
the sources of the  systematic error in \ac{RDM} considered above have been 
completely removed.

There were also some other differences between the 
\ac{RDM} measurements  in both experiments. 
In CMD87 the beam was polarized in the 
VEPP-2M ring itself at the beam energy of about 700~MeV.
%since the booster ring did not yet exist at that time. 
This led to variations 
of the collider parameters before each \ac{RDM} measurement including the
change of the betatron frequencies \( \nu _{x},\, \nu _{z} \) in order 
to pass through intrinsic spin resonances. The imperfection resonance at 
the ``magic energy'' \( E_{beam}=440.65 \) MeV
was crossed adiabatically by decompensating the longitudinal magnetic field
of the detector (so called ``partial siberian snake'' 
mode~\cite{spin_resonance}).
Another parameter affecting the beam energy was the collider 
temperature which changed by approximately 10$^{\circ }$~C
between the  polarization  at the high energy 
and a subsequent \ac{RDM} measurement.

%However, quantitative analysis of the systematics in the two experiments 
%estimates the effects of 
%temperature instability and  changes of the collider parameters for \ac{RDM} 
%measurements as much less than the difference in mass observed
%between the two
%experiments, of the order of 1 MeV. 

In the present experiment the beam
was polarized in the new booster ring at the high energy and after that 
injected into VEPP-2M at the energy of the experiment. Thus, the parameters 
of the collider itself were not changed 
and \ac{RDM} measurements were performed under the same conditions as 
data taking.

The beam energy stability during data taking has been 
thoroughly analysed. This analysis was based both on the deviations of 
about 60 \ac{RDM} measurements at different energies from the predicted 
values and on direct measurements of the beam energy stability by the 
tracking system of the CMD-2 detector \cite{Prep99}. The \ac{RDM}
measurements were consistent with each other in the energy range covering
the \( \omega  \) and \( \phi  \) mesons and showed the  long-term 
beam energy instability of the
order of 50~keV. The latter was used as an uncertainty of the beam energy 
at each point for the calculation of the \( \omega  \)
meson mass systematic error.

%The analysis above describes the reasons for our confidence in the 
%results of this work.  

%Comparing our value of the \( \omega  \) meson mass one can see that it
The value of the  \( \omega  \) meson mass obtained in this work 
is close to the
world average before the CMD87 experiment 
$M_{\omega}=(782.55\pm0.17)$~MeV/\( c^{2} \) (the right shaded bar in 
Fig.~\ref{fig:om_mass}) and is the most precise today. 

\begin{figure}[!htb]
{\centering \begin{tabular}{cc}
\subfigure[$\omega$-meson total width]{\resizebox*{0.49\textwidth}{!}{\includegraphics{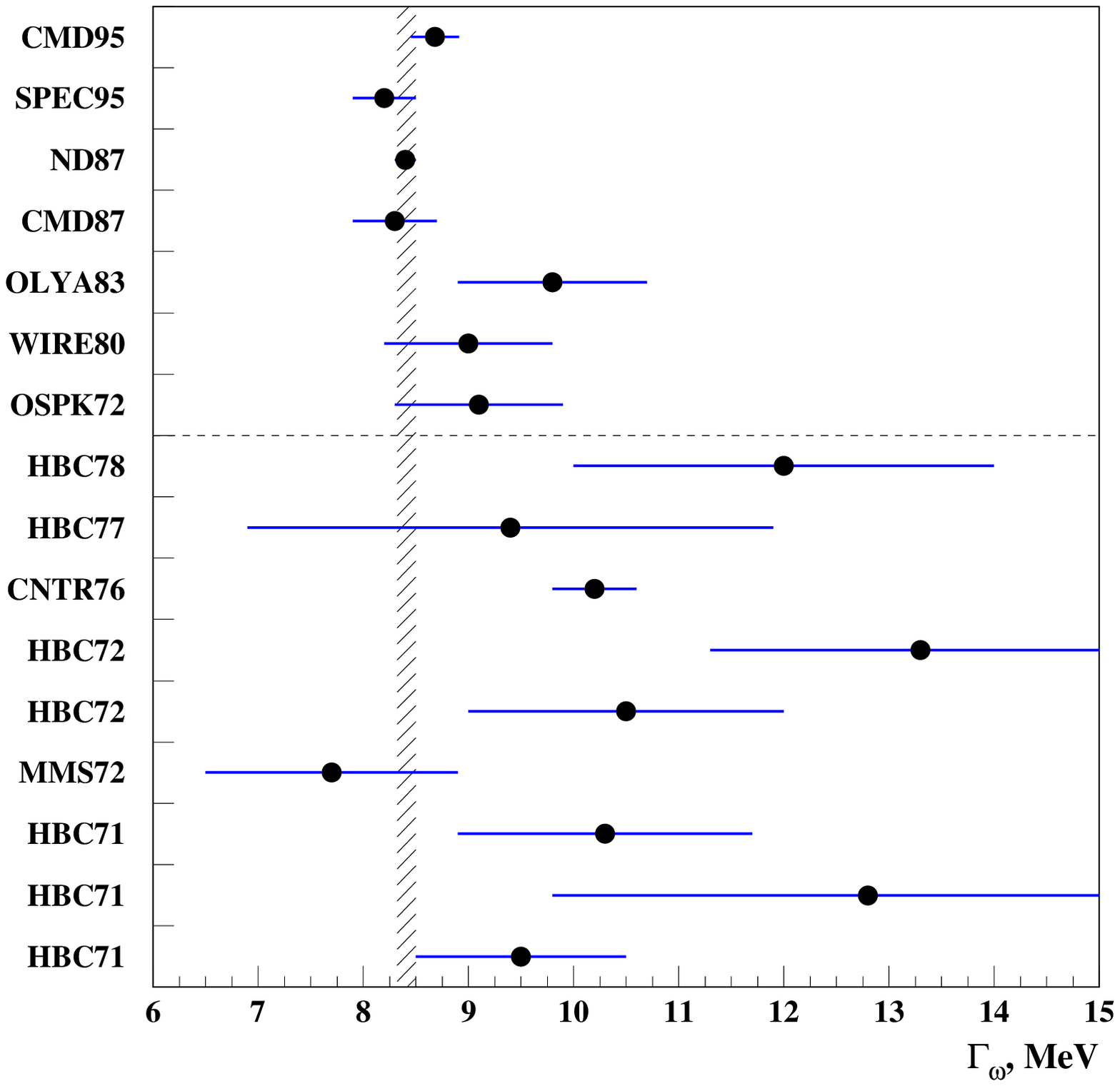}}} &
\subfigure[Leptonic branching ratio
$\Gamma_{\ee}/\Gamma_{\omega}$]{\resizebox*{0.49\textwidth}{!}{\includegraphics{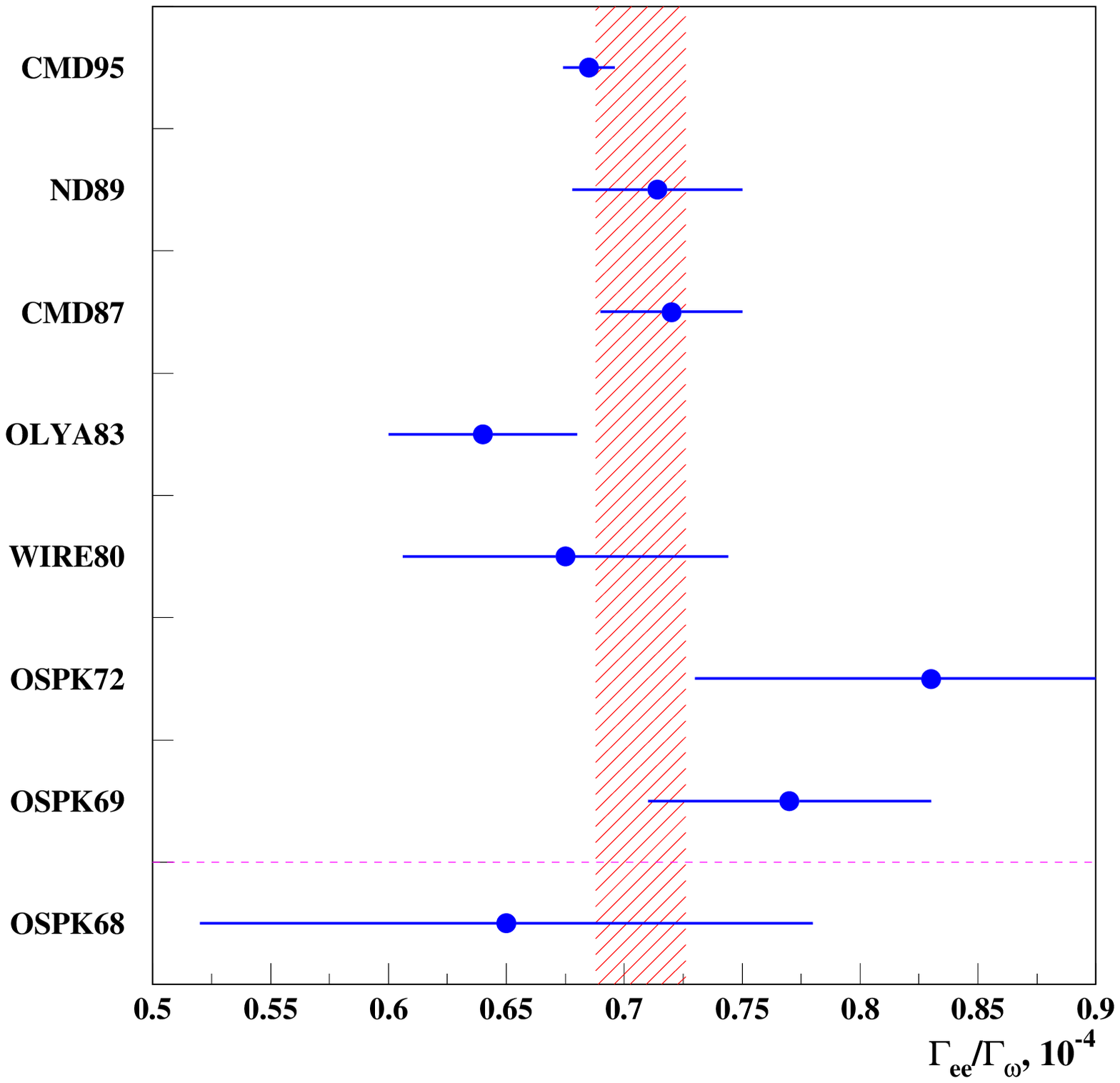}}}
\\
\end{tabular}\par}

\caption{\label{fig:om_width}Experimental data on the 
\protect\( \omega \protect \) meson 
total width~(a) and leptonic branching ratio~(b). Vertical shaded bars 
correspond to the current world averages~\cite{pdg}. The results of this 
work (CMD95) and of experiments below the dashed line are not used 
for averages.}
\end{figure}

The results of this work on the total width of the \( \omega  \) meson 
(Fig.~\ref{fig:om_width}-a)
as well as on the leptonic branching ratio 
\( \Gamma _{\ee }/\Gamma _{\omega } \) (Fig.~\ref{fig:om_width}-b) 
are in good agreement with 
those from  previous experiments. 

\section{Conclusion}

Using the CMD-2 data sample of about 11 200 
\( \omega \rightarrow \pipipi  \) events, the following values of the 
\( \omega  \) meson parameters have been obtained: 
\begin{eqnarray*}
\sigma _{0} & = & (\omcs )\, \mbox {nb},\nonumber \\
M_{\omega } & = & (\ommass )\, \mbox {MeV}/c^{2},\nonumber \\
\Gamma _{\omega } & = & (\omwidth )\, \mbox {MeV},\nonumber \\
\Gamma _{e^{+}e^{-}}\cdot Br(\omega \rightarrow \pipipi ) & = & (\omgeebr )\cdot 10^{-3}\, \mbox {MeV}.\nonumber 
\end{eqnarray*}

These results, except for the total width, are more precise than those from 
previous experiments.
The mass value differs significantly from the previous most precise 
measurement \cite{CMD87} which was performed by a group including many 
authors of this work. Due to the present more thorough study of
systematic errors, our mass measurement supersedes that of Ref. \cite{CMD87}.

\section{Acknowledgements}

The authors are grateful to the staff of VEPP-2M for the excellent 
performance of the collider and to all engineers and technicians 
who participated in the design,
commissioning and operation of CMD-2.

\end{document}